\documentclass[superscriptaddress, twocolumn, prl, aps]{revtex4-1}

\usepackage{graphicx}
\usepackage{amsmath, amsthm, amssymb, bm}

\begin{document}

\title{Scaling Behavior of Anisotropy Relaxation in Deformed Polymers}
\author{Christopher N. Lam}
\thanks{These authors contributed equally to this work.}
\affiliation{Center for Nanophase Materials Sciences, Oak Ridge National Laboratory, Oak Ridge, Tennessee 37831, USA}
\author{Wen-Sheng Xu}
\thanks{These authors contributed equally to this work.}
\affiliation{Center for Nanophase Materials Sciences, Oak Ridge National Laboratory, Oak Ridge, Tennessee 37831, USA}
\author{Wei-Ren Chen}
\affiliation{Neutron Scattering Division, Oak Ridge National Laboratory, Oak Ridge, Tennessee 37831, USA}
\author{Zhe Wang}
\affiliation{Neutron Scattering Division, Oak Ridge National Laboratory, Oak Ridge, Tennessee 37831, USA}
\affiliation{Department of Engineering Physics, Tsinghua University, Beijing 100084, China}
\author{Christopher B. Stanley}
\affiliation{Neutron Scattering Division, Oak Ridge National Laboratory, Oak Ridge, Tennessee 37831, USA}
\author{Jan-Michael Y. Carrillo}
\affiliation{Center for Nanophase Materials Sciences, Oak Ridge National Laboratory, Oak Ridge, Tennessee 37831, USA}
\affiliation{Computational Sciences and Engineering Division, Oak Ridge National Laboratory, Oak Ridge, Tennessee 37831, USA}
\author{David Uhrig}
\affiliation{Center for Nanophase Materials Sciences, Oak Ridge National Laboratory, Oak Ridge, Tennessee 37831, USA}
\author{Weiyu Wang}
\affiliation{Center for Nanophase Materials Sciences, Oak Ridge National Laboratory, Oak Ridge, Tennessee 37831, USA}
\author{Kunlun Hong}
\affiliation{Center for Nanophase Materials Sciences, Oak Ridge National Laboratory, Oak Ridge, Tennessee 37831, USA}
\author{Yun Liu}
\affiliation{Center for Neutron Research, National Institute of Standards and Technology, Gaithersburg, Maryland 20899, USA}
\affiliation{Department of Chemical and Biomolecular Engineering, University of Delaware, Newark, Delaware 19716, USA}
\author{Lionel Porcar}
\affiliation{Institut Laue-Langevin, B.P. 156, F-38042 Grenoble CEDEX 9, France}
\author{Changwoo Do}
\affiliation{Neutron Scattering Division, Oak Ridge National Laboratory, Oak Ridge, Tennessee 37831, USA}
\author{Gregory S. Smith}
\affiliation{Neutron Scattering Division, Oak Ridge National Laboratory, Oak Ridge, Tennessee 37831, USA}
\author{Bobby G. Sumpter}
\affiliation{Center for Nanophase Materials Sciences, Oak Ridge National Laboratory, Oak Ridge, Tennessee 37831, USA}
\affiliation{Computational Sciences and Engineering Division, Oak Ridge National Laboratory, Oak Ridge, Tennessee 37831, USA}
\author{Yangyang Wang}
\email{wangy@ornl.gov}
\affiliation{Center for Nanophase Materials Sciences, Oak Ridge National Laboratory, Oak Ridge, Tennessee 37831, USA}

\begin{abstract}
Drawing an analogy to the paradigm of quasi-elastic neutron scattering, we present a general approach for quantitatively investigating the spatiotemporal dependence of structural anisotropy relaxation in deformed polymers by using small-angle neutron scattering. Experiments and non-equilibrium molecular dynamics simulations on polymer melts over a wide range of molecular weights reveal that their conformational relaxation at relatively high momentum transfer \textit{Q} and short time can be described by a simple scaling law, with the relaxation rate proportional to \textit{Q}. This peculiar scaling behavior, which cannot be derived from the classical Rouse and tube models, is indicative of a surprisingly weak direct influence of entanglement on the microscopic mechanism of single-chain anisotropy relaxation.
\end{abstract}

\date{\today}

\maketitle

Dynamics of polymers are characterized by a remarkably wide range of length and time scales. Historically, the development of quasi-elastic neutron scattering techniques has provided a powerful tool for understanding the microscopic details of polymer motions in the quiescent state, where the spatial and temporal dependence of dynamics is encapsulated in the measured coherent and incoherent dynamic structure factors or corresponding intermediate scattering functions in Fourier space \cite{deGennes1967, DuboisVioletteDeGennes1967, EwenRichter}. The application of quasi-elastic scattering to polymers undergoing mechanical deformation and flow, however, has so far been limited by serious theoretical and practical difficulties, despite some technical progress \cite{Dixon, Pyckhout-Hintzen, rheoXPCS, Kawecki}. In this context, time-resolved small-angle scattering, based on either \textit{ex-situ} or \textit{in-situ} methods, presents an alternative approach to elucidating the microstructure of polymers in the deformed state.

For quasi-elastic scattering, dynamic structure factors and intermediate scattering functions [$F(Q,t)$] provide a convenient mathematical language for describing the spatiotemporal dependence of molecular dynamics \cite{EwenRichter}. As is often the case, the $F(Q,t)$ measured at different momentum transfers $Q$ can be analyzed as stretched exponential functions: $F(Q,t)\sim\exp\lbrace-\left[t/\tau(Q)\right]^{\beta}\rbrace$, and the corresponding relaxation time $\tau(Q)$ can be further described by a power-law function: $\tau(Q)\sim Q^{-\alpha}$. The spatial and temporal dependence of molecular relaxation can thus be quantified through the exponents $\alpha$ and $\beta$. In contrast, a mature framework has not emerged for structural anisotropy relaxation of polymers in non-equilibrium, deformed states. Consequently, in spite of decades of research using small-angle neutron scattering (SANS) and computer simulation, there has been no \textit{quantitative} analysis of the spatiotemporal dependence of molecular relaxation in deformed polymers.

Motivated by this challenge, in this Letter, we describe a quantitative method for analyzing the structural anisotropy relaxation of polymers, by drawing an analogy to the paradigm of quasi-elastic neutron scattering. To illustrate our approach, let us consider the case of small-angle neutron scattering (SANS) from a mixture of two identical polymers, one deuterated and the other protonated, where the scattering intensity $I(\bm{Q};t)$ is dominated by the single-chain structure factor $S(\bm{Q};t)$ \cite{deGennesBook, HigginsBook}:
\begin{equation}
I(\bm{Q};t)\propto S(\bm{Q};t)=\frac{1}{N^2}\sum_{m,n}\left\langle e^{i \bm{Q}\cdot [\bm{R}_n(t) -\bm{R}_m(t)]} \right\rangle.
\end{equation}
Here $\bm{Q}$ is the scattering wavevector, $t$ is time, and $\bm{R}_n$ and $\bm{R}_m$ are the position vectors of the \textit{n}th and \textit{m}th segments of a polymer chain of length $N$, respectively. The notation $f(\dots;t)$ is used to emphasize the fact that we are measuring the \textit{time evolution} of the quantity $f(\dots)$, instead of its \textit{time correlation}. For a step-strain experiment in which the sample is deformed instantaneously at time $t=0$, $S(\bm{Q};t)$ describes the relaxation of the perturbed polymer structure towards the equilibrium state. The approach we introduce here exploits the so-called spherical harmonic expansion technique \cite{ClarkAckerson, EvansHanleyHess, Suzuki1987, WagnerAckerson, WangPRX, Huang1, Huang2}. In general, the single-chain structure factor can be decomposed by a series of spherical harmonics: $S(\bm{Q};t)=\sum_{l,m}S_l^m(Q;t)Y_l^m(\theta,\phi)$, where $\theta$ is the polar angle, $\phi$ is the azimuthal angle, and $S_l^m(Q;t)$ is the expansion coefficient corresponding to each real spherical harmonic function $Y_l^m(\theta,\phi)$. For the aforementioned step-strain relaxation experiment, a class of anisotropic structural relaxation functions $\phi_l^m(Q;t)$ can be defined for all the $l>0$ terms:
\begin{equation}
\phi_l^m(Q;t) \equiv S_l^m(Q;t)/S_l^m(Q;0).
\end{equation}
These functions bear an apparent resemblance to the classical intermediate scattering functions and encapsulate the essential spatial and temporal information about the anisotropic single-chain structure.
\begin{figure}
\centering
\includegraphics[scale=1.0]{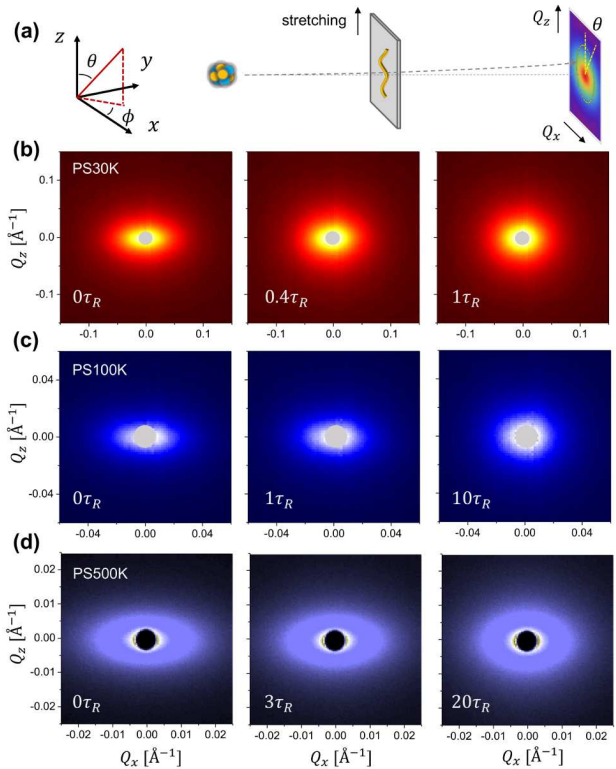}
\caption{(a) Illustration of the configuration of the SANS experiments on the uniaxially deformed polymer films: the stretching is along the \textit{z}-axis whereas the incident SANS beam is perpendicular to the \textit{xz}-plane. (b), (c), and (d) present the SANS spectra during different stages of relaxation for PS30K, PS100K, and PS500K, respectively. Different color schemes are used for these samples. $\tau_R$ stands for the Rouse relaxation time.}
\label{F1}
\end{figure}

\begin{figure}
\centering
\includegraphics[scale=0.52]{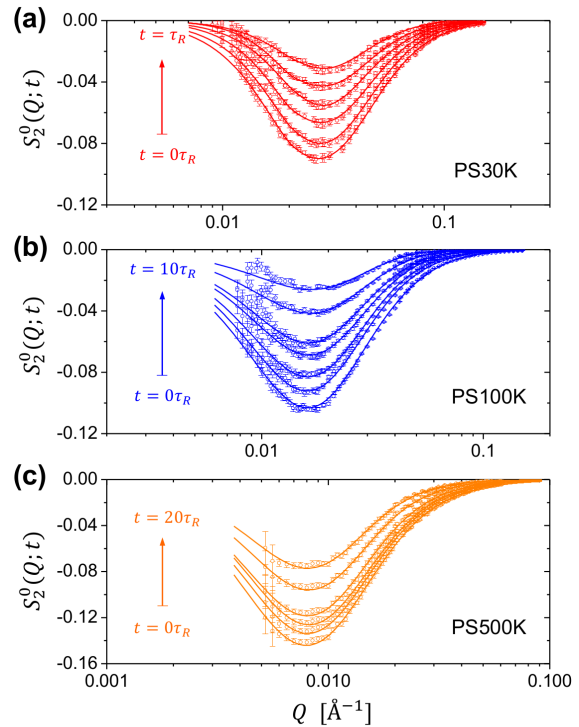}
\caption{Relaxation of the leading anisotropic expansion coefficient $S_2^0(Q)$ after a step deformation. (a) PS30K at 0, 0.1, 0.2, 0.4, 0.6, and 1$\tau_R$. (b) PS100K at 0, 0.5, 1, 2, 3, 6, and 10$\tau_R$. (c) PS500K at 0, 0.5, 1, 3, 10, and 20$\tau_R$. Solid lines: guide to eye.}
\label{F2}
\end{figure}

\begin{figure*}
\centering
\includegraphics[scale=0.54]{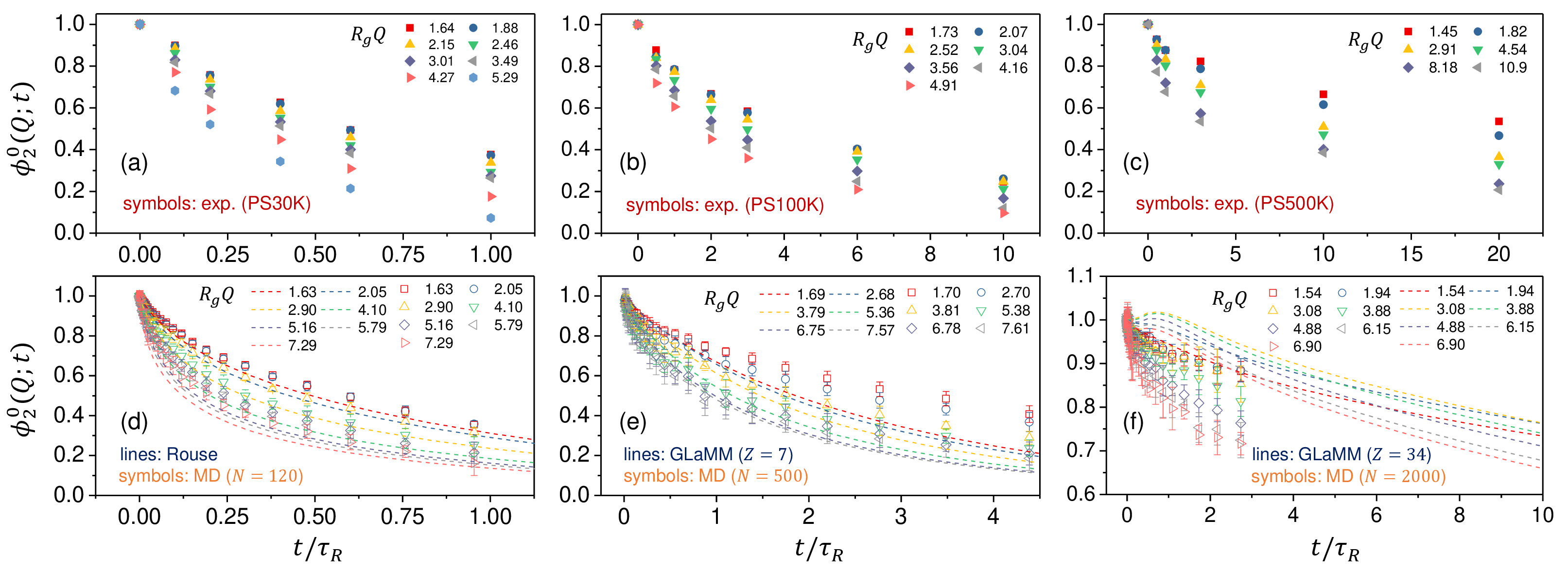}
\caption{Anisotropic structural relaxation function $\phi_2^0(Q;t)$ as a function of the normalized relaxation time $t/\tau_R$ (with $\tau_R$ being the Rouse relaxation time) at different $Q$s for experiments [(a) PS30K, (b) PS100K, and (c) PS500K] and for theories and simulations [(d), (e), and (f)]. To put these results in perspective, the momentum transfer $Q$ is normalized by the equilibrium radius of gyration, $R_g$, of the polymer. The GLaMM calculations include the ``local fluctuation" effect \cite{Blanchard}.}
\label{F3}
\end{figure*}

To demonstrate the usefulness of this new approach, we performed small-angle neutron scattering experiments on a series of uniaxially deformed polymers [Fig. \ref{F1}(a)] at the EQ-SANS beamline at the Oak Ridge National Laboratory and the NGB 30m SANS beamline at the National Institute of Standards and Technology. Our experimental system consisted of isotopically labeled polystyrenes (PS) of three different molecular weights, which we shall refer to as PS30K, PS100K, and PS500K. The details of the compositions of the samples, their molecular characteristics, and the experimental procedures are provided in the Supplemental Material \cite{SM}. The PS samples were uniaxially elongated on an RSA-G2 solid analyzer in their melt state to a stretching ratio $\lambda=1.8$, allowed to relax for different amounts of time after the step deformation, and subsequently quenched to the glassy state by pumping cold air into the environmental test chamber. In our experiments, the time required to effectively freeze the large-scale molecular motions was negligibly small compared to the Rouse or reptation relaxation times, ensuring that the evolution of $S(\bm{Q})$ could be captured with sufficient temporal accuracy.

The SANS spectra from the plane parallel to the stretching direction (the \textit{xz}-plane) become increasingly isotropic as the stress relaxation progresses [Figs. \ref{F1}(b), \ref{F1}(c), and \ref{F1}(d)]. The spherical harmonic expansion technique outlined in our previous work \cite{WangPRX} allows us to transform these two-dimensional SANS spectra into plots of wavenumber-dependent expansion coefficients [Figs. \ref{F2}(a), \ref{F2}(b), and \ref{F2}(c)]. The axial symmetry of the uniaxial deformation eliminates the dependence of $S(\bm{Q};t)$ on $\phi$ and forbids all the $m \neq 0$ and odd $l$ terms, namely:
\begin{equation}
S(\bm{Q};t)=\sum_{l:\mathrm{even}}S_l^0(Q;t)Y_l^0(\theta).
\end{equation}
In this work, we confine ourselves to the analysis of the leading anisotropic expansion coefficient $S_2^0(Q;t)$ and its corresponding relaxation function $\phi_2^0(Q;t) \equiv S_2^0(Q;t)/S_2^0(Q;0)$. In the Supplemental Material \cite{SM}, we show that according to the classical theory the tensile stress of Gaussian chains is determined by the two-point spatial correlations associated with only the real spherical harmonic function $Y_2^0(\theta,\phi)$. Therefore, $\phi_2^0(Q;t)$ contains the relevant information of the structural changes underlying the macroscopic stress relaxation.

The anisotropic structural relaxation function $\phi_2^0(Q;t)$ can be examined by presenting $\phi_2^0(Q;t)$ as a function of the duration of stress relaxation \textit{t} at different scattering wavenumbers $Q$ [Figs. \ref{F3}(a), \ref{F3}(b), and \ref{F3}(c)]. This approach, in its apparent form, is analogous to the classical way in which the intermediate functions of polymers are analyzed \cite{EwenRichter}. Here, we focus on the intermediate- and high-$Q$ regions, i.e., $R_g Q\agt 1$, corresponding to length scales that are roughly equal to or smaller than the size of the polymer coil. Figures \ref{F3}(a), \ref{F3}(b), and \ref{F3}(c) reveal that the anisotropy relaxation depends highly on the length scale probed by the scattering experiment --- the $\phi_2^0$ at high $Q$s relax much faster than those at low $Q$s. This result is in accordance with our current general understanding of liquid dynamics. However, it is worth noting that neither the affine deformation model \cite{Ullman1979, Muller, Hassager}, which is often used to interpret the SANS spectra of deformed polymers, nor the speculative formula proposed by de Gennes \cite{SM, deGennesLeger}, anticipates such wavenumber dependence for the anisotropic structural relaxation. For example, the phenomenological approach by de Gennes and L\'{e}ger predicts that the rate of anisotropy relaxation is independent of $Q$ \cite{SM}. To lend support to our experimental observation, we carried out complementary non-equilibrium molecular dynamics (MD) simulations of the step-strain experiments based on the coarse-grained bead-spring model for polymer melts \cite{KremerGrest, Xu2018}. Three different chain lengths, $N=120$, $500$, and $2000$, were simulated to mirror the PS30K, PS100K, and PS500K samples, respectively. The details of the simulations are described in the Supplemental Material \cite{SM}. Figures \ref{F3}(d), \ref{F3}(e) and \ref{F3}(f) show that the MD simulations produce qualitatively similar behavior for the anisotropic structural relaxation function $\phi_2^0(Q;t)$, which further confirms the experimental results.
\begin{figure}
\centering
\includegraphics[scale=0.54]{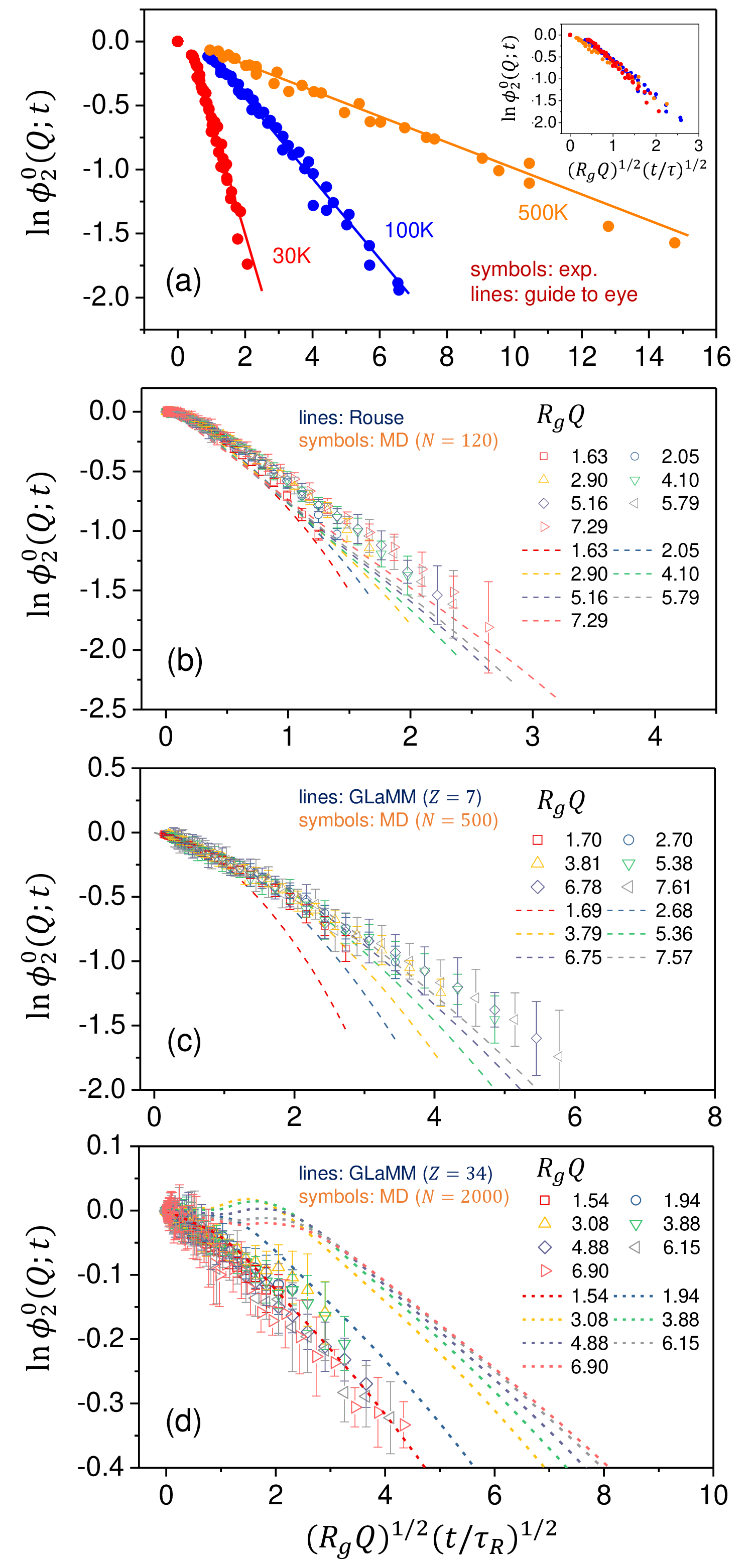}
\caption{$\ln[\phi_2^0(Q;t)]$ as a function of the scaling variable $(R_g Q)^{1/2}(t/\tau_R)^{1/2}$ for the experimental, theoretical, and simulation results presented Fig. 3. Inset of (a) shows $\ln[\phi_2^0(Q;t)]$ as a function of $(R_g Q)^{1/2}(t/\tau)^{1/2}$, where $\tau$ is the longest viscoelastic relaxation time of the sample. }
\label{F4}
\end{figure}

It is intriguing to ask whether the classical Rouse \cite{Rouse} and tube models \cite{deGennes1, DEBook, Graham} can offer some insights into the observed spatiotemporal dependence of anisotropy relaxation. Before embarking on the theoretical analysis, we point out a striking feature of the relaxation behavior of $\phi_2^0(Q;t)$, observed in both the SANS experiments and MD simulations. In the case of quasi-elastic neutron scattering, the so-called Rouse scaling approach, which stems from de Gennes' derivation of the dynamic structure (form) factors of the Rouse model \cite{deGennes1967, EwenRichter, RichterPRL1989}, has been fruitful in elucidating the slow dynamics of polymer melts. In particular, it has been shown that for $R_g Q>1$ and $t<\tau_R$, the coherent intermediate scattering function $F_{\mathrm{coh}}(Q,t)$ can be described as a function of the scaling variable $(\Gamma t)^{1/2}$, where the decay rate $\Gamma \propto Q^4$. Therefore, $F_{\mathrm{coh}}(Q,t)$ measured at different $Q$ and $t$ can be collapsed by plotting $F_{\mathrm{coh}}(Q,t)$ against the Rouse variable $(\Gamma t)^{1/2}$ \cite{RichterPRL1989}.  Interestingly, we find that the anisotropic structural relaxation functions $\phi_2^0(Q;t)$ at different $Q$s can be reduced to a single curve by using $(R_g Q)^{1/2}(t/\tau_R)^{1/2}$ as the scaling variable [Fig. \ref{F4}(a)]. In other words, Fig. \ref{F4}(a) indicates that $\phi_2^0(Q;t)$ can be cast into the following functional form:
\begin{equation}
\frac{S_2^0(Q;t)}{S_2^0(Q;0)}=\phi_2^0(Q;t)\sim \exp \left[-(\Gamma t)^{1/2}\right],
\end{equation}
with $\Gamma \propto Q$. Furthermore, it appears that the data from the three systems can be further superimposed onto a master curve by normalizing the time $t$ with the longest viscoelastic relaxation time $\tau$ [inset of Fig. \ref{F4}(a)]. Similar to the case of quasi-elastic neutron scattering, the above scaling approach is only valid under roughly the condition of $R_g Q >1$ and $t<\tau$. Lastly, the experimentally observed scaling behavior is corroborated by the non-equilibrium MD simulations [Figs. \ref{F4}(b), \ref{F4}(c), and \ref{F4}(d)]. 

To put these results in perspective, we evaluate the anisotropic structural relaxation function $\phi_2^0(Q;t)$ using the Rouse model for the unentangled system and the tube model by Graham et al. \cite{Graham}, i.e. the GLaMM model, for the entangled polymers \cite{SM}. The comparisons between the theories and simulations are presented in Figs. \ref{F3}(d), \ref{F3}(e), \ref{F3}(f), \ref{F4}(b), \ref{F4}(c), and \ref{F4}(d). At first glance, the Rouse model seems to be able to provide a reasonable description of the spatiotempoeral dependence of anisotropy relaxation for short chains [Fig. \ref{F3}(d)]. However, a closer inspection reveals that the model does not faithfully reproduce the aforementioned scaling for $\phi_2^0(Q;t)$ [Fig. \ref{F4}(b)]. Additionally, it should be noted that for the entangled chains, the scaling behavior persists well beyond the Rouse relaxation time. The observed scaling behavior for anisotropy relaxation, therefore, does not arise from the unconstrained Rouse motion. It is worth mentioning that deviation from the standard Rouse behavior has been observed in neutron spin-echo experiments on an unentangled polyethylene melt in the equilibrium state \cite{PaulPRL1998} and is attributed to non-Gaussian dynamics. The possible connection between these phenomena remains to be explored. 

Figures \ref{F3}(e), \ref{F3}(f), \ref{F4}(c), and \ref{F4}(d) indicate that the tube model also fails to predict the correct spatiotemporal dependence for anisotropy relaxation, even with the consideration of local fluctuations about the primitive path \cite{SM, Blanchard}. We previously showed that the chain retraction mechanism of the tube model leads to an increase of $S_2^0(Q;t)$ around the Rouse time in the intermedate $Q$ range after a large step uniaxial deformation, which is inconsistent with both the SANS experiment \cite{WangPRX} and the MD simulation \cite{Xu2018}. Not surprisingly, quantitative analysis of $\phi_2^0(Q;t)$ reveals a strong deviation from the theoretical prediction for the well-entangled systems (PS500K and the $N=2000$ chain in simulation). The effect of chain retraction is less pronounced for mildly entangled polymers \cite{Xu2018}. Nevertheless, the theory disagrees with both the experiment [Fig. \ref{F4}(a)] and the simulation [Fig. \ref{F4}(c)].

The most surprising aspect of our result is that entanglement appears to have a very weak direct influence on the microscopic mechanism of single-chain anisotropy relaxation: the same peculiar behavior is observed for both entangled and unentangled systems. Furthermore, for the well-entangled systems, the scaling law holds well both below ($aQ>1$) and above ($aQ<1$) the length scale of the tube diameter $a$. The effect of entanglement shows up only indirectly through the scaling variable $\tau$ [inset of Fig. \ref{F4}(a)]. This observation suggests a possible simple explanation for the absence of ``chain retraction" in the previous step-strain relaxation experiments \citep{WangPRX} and molecular dynamics simulations \cite{Xu2018}: if the confinement effect of the tube on the test chain is much weaker than what it is supposed to be, then the molecular relaxation on the time scale of the Rouse time will not produce the unique conformation predicted by the chain retraction mechanism of the tube model \cite{DEBook}. 

In summary, we present the first quantitative analysis of the spatial and temporal dependence of anisotropy relaxation in deformed polymers by using small-angle neutron scattering and non-equilibrium molecular dynamics simulations. We show that the relaxation of the anisotropic structure of   uniaxially stretched entangled and unentangled polymer melts can be described by a simple, universal scaling law, with the relaxation rate proportional to the magnitude of the momentum transfer. This unexpected finding presents a challenge to our current theoretical understanding of the rheological behavior of polymers and stresses the importance of studying the spatiotemporal dependence of molecular motion under deformation and flow --- an aspect that has been overlooked in traditional polymer rheology.

\begin{acknowledgements}
This research was sponsored by the Laboratory Directed Research and Development Program of Oak Ridge National Laboratory, managed by UT Battelle, LLC, for the U.S. Department of Energy. W.-R. Chen and Z. Wang acknowledge the support by the U.S. Department of Energy, Office of Science, Office of Basic Energy Sciences, Materials Sciences and Engineering Division. The material characterization and SANS experiments were performed at the Center for Nanophase Materials Sciences and the EQ-SANS beamline of the Spallation Neutron Source, respectively, which are DOE Office of Science User Facilities. This research used resources of the Oak Ridge Leadership Computing Facility at the Oak Ridge National Laboratory, which is supported by the Office of Science of the U.S. Department of Energy under Contract No. DE-AC05-00OR22725. Access to NGB 30m SANS was provided by the Center for High Resolution Neutron Scattering, a partnership between the National Institute of Standards and Technology and the National Science Foundation under Agreement No. DMR-1508249.
\end{acknowledgements}

\end{document}